\documentclass{aa} 

\usepackage{natbib}
\usepackage[binary-units=true]{siunitx}
\bibpunct{(}{)}{;}{a}{}{,}

\usepackage{graphicx}
\usepackage{txfonts}

\usepackage{soul}
\usepackage{multirow}
\usepackage{footnote}
\usepackage{verbatim}

\makesavenoteenv{tabular}
\makesavenoteenv{table}

\usepackage{caption}
\usepackage{subcaption}

\usepackage{tablefootnote}
\usepackage{threeparttable}

\def\arcsec{$^{\prime\prime}$}
\def\arcmin{$^{\prime}$}

\begin{document}

   \title{First observations from the SPICE EUV spectrometer on Solar Orbiter}

\titlerunning{First observations from SPICE}


   \author{A.~Fludra\inst{\ref{i:ral}}\fnmsep\thanks{Corresponding author: Andrzej Fludra \email{andrzej.fludra@stfc.ac.uk}}
        \and
         M.~Caldwell\inst{\ref{i:ral}}
        \and
         A.~Giunta\inst{\ref{i:ral}}
        \and
         T.~Grundy\inst{\ref{i:ral}}
        \and
         S.~Guest\inst{\ref{i:ral}}
        \and
         S.~Leeks\inst{\ref{i:ral}}
        \and
         S.~Sidher\inst{\ref{i:ral}}
        \and
         F.~Auch\`ere\inst{\ref{i:ias}}
        \and
         M.~Carlsson\inst{\ref{i:uio}}
        \and
         D.~Hassler\inst{\ref{i:swri}}
        \and
         H.~Peter\inst{\ref{i:mps}}
        \and
         R.~Aznar Cuadrado\inst{\ref{i:mps}}
        \and 
         É.~Buchlin\inst{\ref{i:ias}}
        \and 
         S.~Caminade\inst{\ref{i:ias}}
        \and
         C.~DeForest\inst{\ref{i:swri}}
        \and
         T.~Fredvik\inst{\ref{i:uio}}
        \and
         M.~Haberreiter\inst{\ref{i:pmod}}
        \and
         L.~Harra\inst{\ref{i:pmod},\ref{i:eth}}
        \and
         M.~Janvier\inst{\ref{i:ias}}
        \and
         T.~Kucera\inst{\ref{i:nasa-gsfc}}
        \and
         D.~M\"uller\inst{\ref{i:estec}}
        \and
         S.~Parenti\inst{\ref{i:ias}}
        \and
         W.~Schmutz\inst{\ref{i:pmod}}
        \and
         U.~Sch\"uhle\inst{\ref{i:mps}} 
        \and
         S.K.~Solanki\inst{\ref{i:mps},\ref{i:ssr}}
        \and
         L.~Teriaca\inst{\ref{i:mps}}
        \and 
         W.T.~Thompson\inst{\ref{i:adnet}}  
        \and
         S.~Tustain\inst{\ref{i:ral}}
        \and
         D.~Williams\inst{\ref{i:esac}}
        \and
         P.R.~Young\inst{\ref{i:nasa-gsfc},\ref{i:nuu}}
        \and
         L.P.~Chitta\inst{\ref{i:mps}} 
    }
    \institute{
            RAL Space, UKRI STFC Rutherford Appleton Laboratory, Didcot, United Kingdom\label{i:ral}
           \and
           Université Paris-Saclay, CNRS, Institut d’Astrophysique Spatiale, 91405, Orsay, France \label{i:ias}
            \and
            Institute of Theoretical Astrophysics, University of Oslo, Norway\label{i:uio}
            \and
            Southwest Research Institute, Boulder, CO, USA\label{i:swri}
            \and
            Max-Planck-Institut für Sonnensystemforschung, G\"ottingen, Germany\label{i:mps}
            \and
            Physikalisch-Meteorologisches Observatorium Davos, World Radiation Center,  Davos Dorf, Switzerland\label{i:pmod}
            \and
            NASA Goddard Space Flight Center, Greenbelt, MD, USA\label{i:nasa-gsfc}
            \and
            European Space Agency, ESTEC, Noordwijk, The Netherlands\label{i:estec}
            \and
            ADNET Systems Inc., NASA Goddard Space Flight Center, Greenbelt, MD, USA\label{i:adnet}
            \and
            European Space Agency, ESAC, Villanueva de la Cañada, Spain\label{i:esac}
            \and
            ETH Z\"urich, IPA, HIT building, Wolfgang-Pauli-Str. 27, 8093 Z\"urich, Switzerland\label{i:eth}
            \and
            School of Space Research, Kyung Hee University, Yongin, Gyeonggi-Do,446-701, Republic of Korea\label{i:ssr}
            \and
             Northumbria University, Newcastle upon Tyne, NE1 8ST, UK\label{i:nuu}
     }
   
   \date{Received 30 April 2021; accepted 6 September 2021}
   
 
  \abstract
  {}
   {We present first science observations taken during the commissioning activities of the Spectral Imaging of the Coronal Environment (SPICE) instrument on the ESA/NASA Solar Orbiter mission. SPICE is a high-resolution imaging spectrometer operating at extreme ultraviolet (EUV) wavelengths. In this paper we illustrate the possible types of observations to give prospective users a better understanding of the science capabilities of SPICE.  
   }
   {We have reviewed the data obtained by SPICE between April and June 2020 and selected representative results obtained with different slits and a range of exposure times between 5~s and 180~s. Standard instrumental corrections have been applied to the raw data.
   }
   {The paper discusses the first observations of the Sun on different targets and presents an example of the full spectra from the quiet Sun, identifying over 40 spectral lines from neutral hydrogen and ions of carbon, oxygen, nitrogen, neon, sulphur, magnesium, and iron. These lines cover the temperature range between 20,000 K and 1 million K (10MK in flares), providing slices of the Sun’s atmosphere in narrow temperature intervals. We provide a list of count rates for the 23 brightest spectral lines. We show examples of raster images of the quiet Sun in several strong transition region lines, where we have found unusually bright, compact structures in the quiet Sun network, with extreme intensities up to 25 times greater than the average intensity across the image. The lifetimes of these structures can exceed 2.5~hours. We identify them as a transition region signature of coronal bright points and compare their areas and intensity enhancements.  We also show the first above-limb measurements with SPICE above the polar limb in C~III, O~VI, and Ne~VIII lines, and far off limb measurements in the equatorial plane in Mg~IX, Ne~VIII, and O~VI lines. We discuss the potential to use abundance diagnostics methods to study the variability of the elemental composition that can be compared with in situ measurements to help confirm the magnetic connection between the spacecraft location and the Sun’s surface, and locate the sources of the solar wind.
   }
   {The SPICE instrument successfully performs measurements of EUV spectra and raster images that will make vital contributions to the scientific success of the Solar Orbiter mission. }


\maketitle

%

\section{Introduction}
The solar spectrum from 17 to 160 nm contains a huge number of emission lines from species that form at temperatures from 20,000 K to 20 MK in the chromosphere, transition region, and corona. Spectra are critical for determining the plasma physical and chemical characteristics of the emitting source, such as the temperature and density using line ratio or emission measure diagnostics techniques \citep{Del-Zanna:2018}, and for measuring line-of-sight velocities and non-thermal broadening.

Imaging slit spectrometers have been key instruments for the advances in solar physics over the past 25 years, beginning with the Solar Ultraviolet Measurements of Emitted Radiation \citep[SUMER;][]{Wilhelm:1995}, the Ultraviolet Coronagraph Spectrometer \citep[UVCS;][]{Kohl:1995}, and the Coronal Diagnostic Spectrometer \citep[CDS;][]{Harrison:1995} instruments on board the Solar and Heliospheric Observatory (SOHO), launched in 1995. The Extreme Ultraviolet Imaging Spectrometer \citep[EIS;][]{Culhane:2007} was launched on the Hinode spacecraft in 2006 and continues to observe the Sun in the two wavelength bands 17.0–21.2 and 24.6–29.2 nm. These wavebands are dominated by coronal emission lines and also lines from the upper transition region in the 0.1-0.8 MK range \citep{Young:2007}. More recently, the Interface Region Imaging Spectrograph \citep[IRIS;][]{DePontieu:2014} was launched in 2013 and obtains spectra in the three wavelength bands 133.2–135.8~nm, 138.9–140.7~nm and 278.3–283.4~nm which are dominated by chromospheric lines but they also contain the transition region ions Si~IV and O~IV, formed at 80,000 K and 140,000 K, respectively, and a flare line from Fe~XXI, formed at 11 MK.

Despite the great success of these missions, several science questions still remain unsolved. They include the sources and acceleration of the solar wind, the mechanism heating the coronal loops in the quiet Sun and active regions, and the understanding of abundance variations in the transition region and coronal structures, dependent on the first ionisation potential (FIP), and their link with the slow and fast solar wind composition.

The Spectral Imaging of Coronal Environment \citep[SPICE;][]{Spice-all:2020} instrument was launched on Solar Orbiter \citep{Garcia:2021,Mueller:2020} in February 2020. It is a high-resolution imaging spectrometer that observes the Sun in two extreme ultraviolet (EUV) wavelength bands, 70.4–79.0~nm and 97.3-104.9~nm, providing a good balance between cool and hot plasma coverage, complementing the IRIS and EIS instruments. SPICE is capable of recording full spectra in these bands with exposures as short as 1~s, can measure spectra from the disk and low corona, and records all spectral lines simultaneously, using one of three narrow slits: 2{\arcsec}x11{\arcmin}, 4{\arcsec}x11{\arcmin}, 6{\arcsec}x11{\arcmin}, or a wide slit 30{\arcsec}x14{\arcmin}. 
The telescope mirror can be rotated in a direction perpendicular to the slit, normally in small angular steps equal to the slit width.  This allows the spectrometer to record 1D images at adjacent positions on the Sun, and to build 2D raster images of up to 16\arcmin\ in size.

SPICE offers strong emission lines from the chromosphere (20,000 K) to the upper transition region (0.6 MK), as well as coronal lines at temperatures up to 10 MK (see Section~\ref{section:fullspectra}). These lines provide a very good temperature coverage in a single exposure, which is unique among the imaging slit spectrometers flown.

A major science goal of the Solar Orbiter mission \citep{Mueller:2020} is to identify connections between the coronal plasma and the solar wind plasma sampled in~situ by the spacecraft in order to address the above unsolved issues. SPICE is uniquely capable amongst the scientific instruments of remotely measuring the line-of-sight velocity, the temperature, and the composition of the solar wind source regions in the corona. 

The aim of this paper is to present an overview of the first science data taken by SPICE during the commissioning activities. We give examples of different types of observations to illustrate the science capabilities of SPICE and give prospective users a better understanding of potential future applications of SPICE during the forthcoming nominal mission phase. We highlight new observations of very bright sources in the transition region on disk and the ability of SPICE to detect diagnostically useful lines in the corona above the limb. Section~2 describes the spectra and observations of different targets. Section~\ref{section:discussion} presents conclusions and future observations of SPICE. 

\section{Data and results}
\subsection{The commissioning of SPICE}
 SPICE was switched-on on 24 February 2020, and individual subsystems were tested during the following two weeks, including initial tests with cold detectors in dark conditions on 9 March.  All hardware was found to be fully-functional, and showed behaviour consistent with measurements on the ground.  The detector door was opened on 18 March 2020 to allow the detectors to outgas internally within the instrument, while the instrument outer door remained closed for a longer period. This was to protect the optics from receiving any contaminants outgassed from the spacecraft,
which in combination with direct sun-light could cause a loss of EUV throughput.

SPICE is equipped with two intensified detectors consisting of a Microchannel Plate (MCP) intensifier coupled with an Active Pixel Sensor (APS). 
The first tests with the detector high voltage systems were conducted in early April 2020, and the pre-launch baseline for the MCP was set to 900~V in the Short Wavelength (SW) channel, and 850~V in the Long Wavelength (LW) channel. This is to allow a higher gain for measuring the generally weaker lines in the SW channel, while avoiding saturation for the brightest lines in the LW channel (see Section~\ref{section:fullspectra}).  

Following successful opening of the outer doors, SPICE obtained its first light on 21 April 2020. The first SPICE data were taken between 21 April 2020 and 21 June 2020, including the first Solar Orbiter perihelion. The first datasets consisted of a variety of calibration measurements, including some simultaneous observations with other remote sensing Solar Orbiter instruments.  Although the primary goal of this phase was calibration, some initial science measurements were also obtained. Data used in this paper and processing steps are listed in Appendix \ref{appendix:dp}.

The spacecraft distance to the Sun was decreasing from 0.9 to 0.52 AU in this period, reaching the perihelion of 0.52 AU on 16 June 2020.  The data taken at this time are the closest images of the Sun obtained by a spectrometer so far. The size of the SPICE 2\arcsec\ slit was equivalent to 750~km on the Sun.  From early flight data (see Appendix~B), the FWHM of the spatial Point Spread Function (PSF) of SPICE is 6.3 pixel (6.7\arcsec) which is ~2,500~km on the Sun at the first perihelion.

\subsection{Full spectra from SPICE and temperature coverage}
\label{section:fullspectra}
\begin{figure*}
\centering
\includegraphics[width=1.0\hsize]{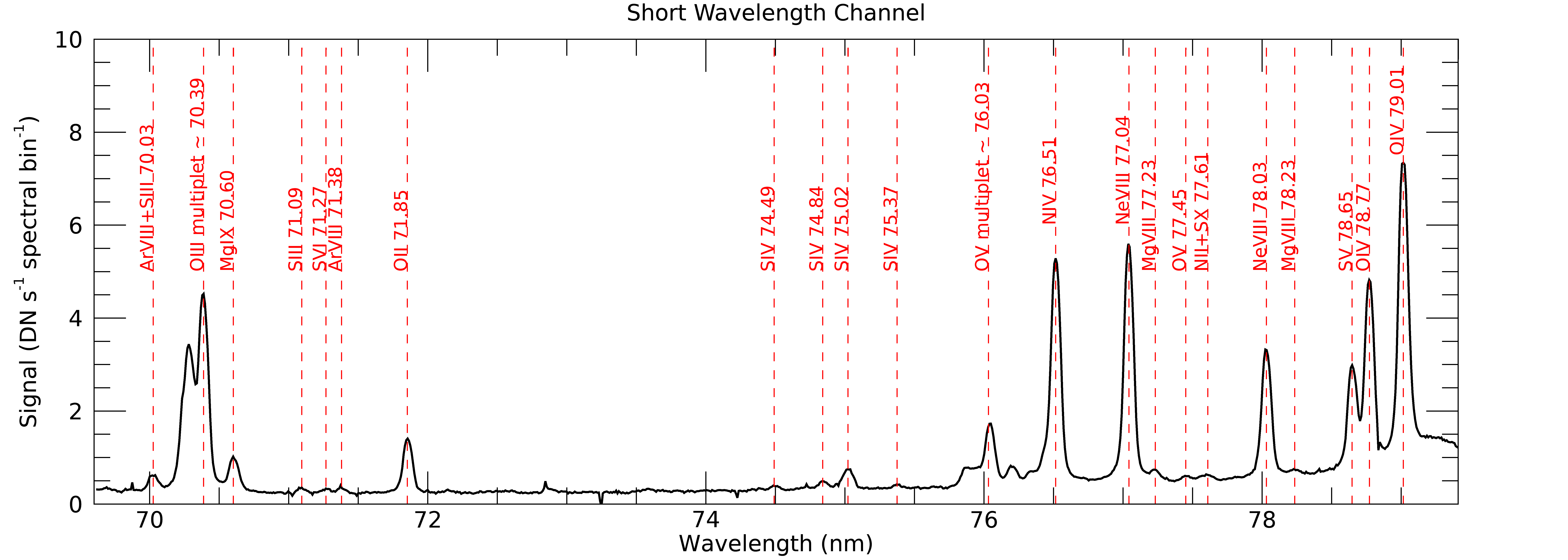}
\includegraphics[width=1.0\hsize]{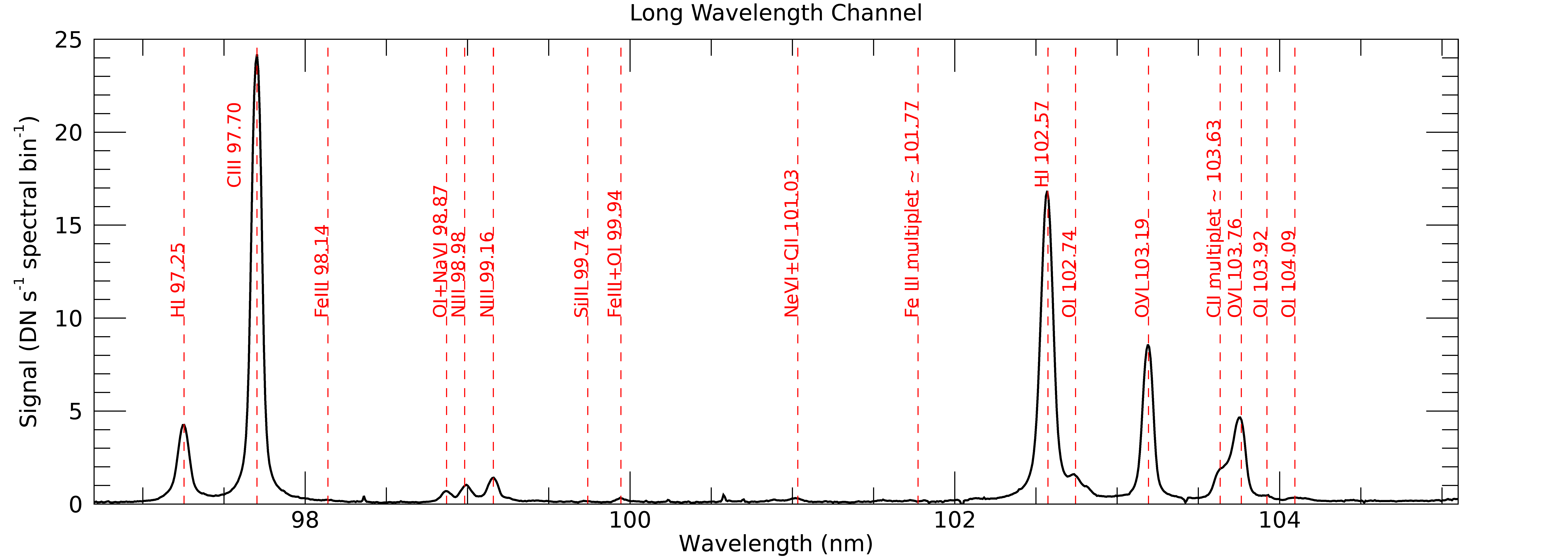}
\caption{Spectra in the two spectral bands of SPICE taken near Sun centre on 21 April 2020, averaged over 400 pixels along the slit. Top panel -- the Short Wavelength band, bottom panel –- the Long Wavelength band.}
\label{fig:1}
\end{figure*}

At the extreme ultraviolet wavelengths in the two SPICE bands, the spectral lines arise from neutral hydrogen and a wide range of ions of carbon, neon, oxygen, nitrogen, sulphur, magnesium, and iron. Figure \ref{fig:1} shows the first spectrum from SPICE taken from a quiet Sun area near disk centre on 21 April 2020, using the 2\arcsec\ slit and a long exposure time of 3 minutes. The spectrum has been averaged over 400 pixels along the slit to increase the signal-to-noise ratio for the faintest lines. It is quite typical of what we can expect from the quiet Sun.

The quiet Sun spectra in the two bands of SPICE, 70.4 - 79.0~nm (SW) and 97.3 - 104.9~nm (LW, the precise limits in both channels change slightly with the instrument’s temperature), contain about two hundred spectral lines. Over 40 of these lines that stand out above the background are labelled in Figure~\ref{fig:1}.

The observed line width (Full Width at Half Maximum, FWHM) is estimated of the order of 8 spectral pixels for the SW band and 9 spectral pixels for the LW band using the 2\arcsec\ slit (see Appendix \ref{appendix:performance}). This corresponds to about 0.07~nm or 200~km~s$^{-1}$ in Doppler units. This means that some lines may result in blends. However, we expect that most of such lines can be separated using a multi-Gaussian fit.

Table~\ref{tab:1} provides a list of 23 most prominent lines from Figure \ref{fig:1}.   The brightest lines (with line peak above 9 DN~s$^{-1}$) are in the Long Wavelength band, in particular C~III 97.70~nm, H~I~Ly-beta 102.57~nm, and O~VI 103.19~nm. They can be measured in the quiet Sun with Signal-to-Noise in the line peak $> 3$ using 5~s exposures. The rest of the lines have much lower intensities. The last column of Table~\ref{tab:1} gives the SPICE Data Numbers (DN/s)  (i.e. uncalibrated count rates on the detector integrated over the line profile.)   

To go one step further and obtain a rough direct comparison between the SW band and the LW band (i.e. a rough relative calibration), the SW DN values can be divided by a factor of 3.4 which is the relative average sensitivity ratio between the two bands. This factor takes into account: (a) the difference in effective area \citep[see][Figure 24]{Spice-all:2020}; (b) the equivalent optical size of the detector pixels for the two channels; (c) the operation of the two detectors at different MCP voltages (SW gain is higher). We note that a full assessment of in-flight performance and calibration will be published at a later date, and this is only intended to account for the design parameters.  

The background seen in SPICE quiet Sun spectra consists of the H~Lyman continuum and C~I continuum as calculated and illustrated using SOHO/SUMER spectra by \cite{Kretzschmar:2004}. We also expect possible instrumental contributions from: (a) electronic background (removed by dark subtraction); (b) in-band scatter (e.g. C~III light scattered outside the profile), normally seen close to the line, for example, apparent broadening of wings; (c) out-of-band scatter (non-SPICE wavelengths, for example, Lyman~$\alpha$ uniformly scattered into the SPICE wavelength range on the detector); (d) out-of-field stray-light (solar radiation from outside of the normal FOV, scattering inside).  The contributions from (b), (c) and (d) have not been quantitatively assessed yet in the in-flight data. 

\begin{table}[]
    \centering
    \caption{Most prominent SPICE lines in the quiet Sun spectrum sorted by increasing wavelength.} 
    \begin{threeparttable}[t]
    \begin{tabular}{|c|r|r|r|}
        \hline
        Ion & $\lambda$[nm] & logT [K] & DN/s \\
        \hline
        O III         &  70.23 & 4.91 &   30.2 \\
        O III         &  70.38 & 4.91 &   40.2 \\
        Mg IX         &  70.60 & 5.96 &   6.8 \\
        O II          &  71.85 & 4.66 &   9.7 \\
        S IV          &  75.02 & 5.01 &   3.6 \\
        O V           &  76.04 & 5.33 &   10.5 \\
        N IV          &  76.52 & 5.08 &   40.5 \\
        Ne VIII       &  77.04 & 5.74 &   43.4 \\
        Mg VIII       &  77.27 & 5.89 &   1.5 \\
        Ne VIII       &  78.03 & 5.74 &   23.7 \\
        S V           &  78.65 & 5.15 &   17.5 \\
        O IV          &  78.77 & 5.15 &   32.0 \\
                      &        &      &        \\
        H Ly $\gamma$ &  97.25 & 4.00 &  39.9 \\
        C III         &  97.70 & 4.81 & 211.6 \\
        N III         &  98.98 & 4.82 &   5.6 \\
        N III         &  99.16 & 4.82 &   12.6 \\
        Ne VI         & 101.03 & 5.56 &   1.2 \\
        H Ly $\beta$  & 102.57 & 4.00 &  171.2 \\
        O I           & 102.74 & 4.21 &   8.8 \\
        O VI          & 103.19 & 5.41 &  76.0 \\
        C II          & 103.60 & 4.47 &  13.1 \\
        C II          & 103.70 & 4.47 &  14.4 \\
        O VI          & 103.76 & 5.41 &  40.6 \\
        \hline
    \end{tabular}
    \begin{tablenotes}
   \item[] The column logT gives the line formation temperature (at the peak of the contribution function from Figure~2). The column DN/s gives uncalibrated count rates from SPICE, obtained from fitting the line profiles of the spectrum in Figure \ref{fig:1}. Optionally, to obtain a rough relative calibration between the SW and LW count rates, the SW band values can be divided by a factor of 3.4 (see text for details).
\end{tablenotes}
\end{threeparttable}
    \label{tab:1}
\end{table}

\begin{figure}
\centering
\includegraphics[width=1.0\hsize]{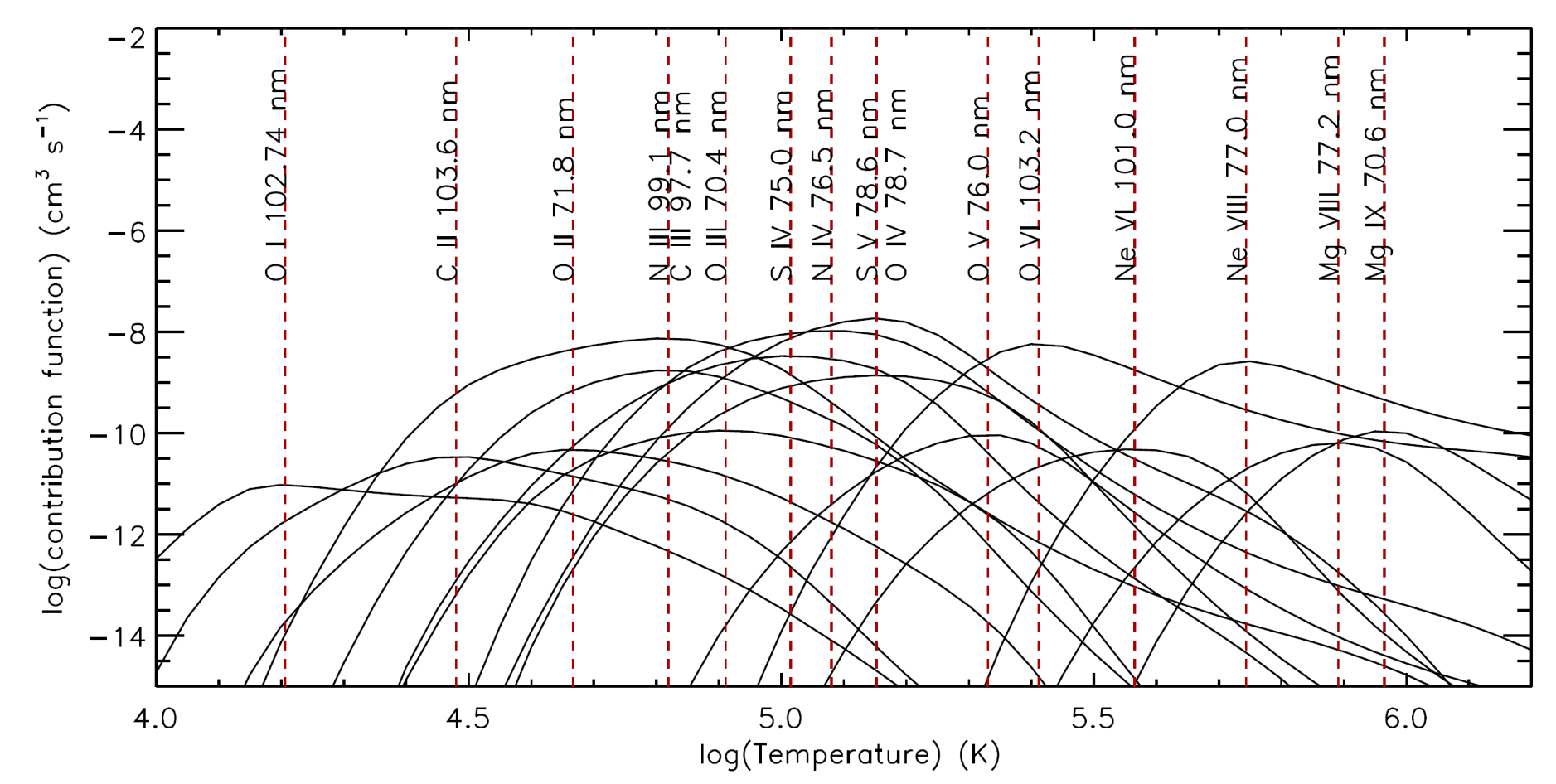}
\caption{Line contribution functions as a function of electron temperature for the lines listed in Table~\ref{tab:1} in the quiet Sun SPICE spectra.   Only one line for each ionisation stage has been included in the plot. Contribution functions have been calculated using ADAS \citep{Summers:2006}.}
\label{fig:2}
\end{figure}

Figure~\ref{fig:2} shows the temperature coverage of SPICE. These curves are the contribution functions of the lines listed in Table~\ref{tab:1}. They are calculated for an optically thin plasma using the Atomic Data and Analysis Structure\footnote{www.adas.ac.uk; open.adas.ac.uk} \citep[ADAS;][]{Summers:2006} set of codes and atomic database, to take into account the possible density effects for all the lines emitted in the upper chromosphere and lower transition region. A constant pressure of $5\times 10^{14}$ K cm$^{-3}$ is used to compute the temperature of the line peaks. The two H~I Lyman lines are not included in the calculation, as they are mostly optically thick and so cannot be treated with an optically thin plasma model.

The lowest temperature lines in the SPICE spectrum are H~Ly~$\beta$, H~Ly~$\gamma$ and O~I emitted from the chromosphere and low transition region. Then we have several transition region lines between 40,000 and 600,000~K. Our best coronal line is Mg~IX 70.60 nm at 910,000 K but it has relatively low intensity, requiring longer exposure times (greater than 90 s, unless binning is used) in the quiet Sun.  In hot active regions and flares we should also see two lines of iron (Fe~XVIII 97.48~nm and Fe~XX 72.15~nm, emitted at 7~MK and 10~MK, respectively).

\subsection{Images from SPICE}
SPICE can also routinely produce rastered images, as explained in Section~1. The scan direction on the disk is from right to left, (i.e. from solar west to solar east, with north at the `top'), although it should be noted that Solar Orbiter is capable of rolling to a different orientation for specific observing campaigns \citep{Auchere:2020}. At each scan position, every pixel along the slit can be acquired simultaneously, with two main spectral options: (1) acquire the full spectrum, covering the full wavelength range; (2) acquire limited wavelength ranges only, up to 8 line profiles of 32-spectral bins wide.

Then, by selecting any spectral line profile from the recorded spectrum, we can build an image of summed intensities in that line. 
As a further option, the wavelength coverage can be increased in the second case by summing together adjacent spectral bins on-board, in groups of 2 or 4 pixels. This option is currently under evaluation for effectiveness in future measurements. 

The SPICE instrument design is optimised to make efficient use of spectral line profiles for large or fast rasters. The instrument is capable of performing all on-board processing of data from the line profiles (including compression) while the raster scan is in progress, and efficiency is limited only by the operation of the scan mirror mechanism.  With overheads of 0.25~s per scan step, and 10~s between repeats of a full raster scan, the observation time is usually dominated by the exposure time required.  Typical examples are:
\begin{itemize}
    \item Dynamics studies: 5~s exposures, 1.1\arcmin\ FOV, 2\arcsec\ steps: 2m58s cadence (18~s overhead).
    \item Large FOV Synoptic: 20~s exposures, 15\arcmin\ FOV, 4\arcsec\ steps: 1h15m36s cadence (66~s overhead).
    \item Composition Map: 90~s exposures, 8.5\arcmin\ FOV, 4\arcsec\ steps: 3h12m42s cadence (42~s overhead).
    \item Full Spectrum: 60~s exposures, 2\arcmin\ FOV, 4\arcsec\ steps: 
3150~s cadence (1350~s overhead)
\end{itemize}

We note that the full spectrum option requires more telemetry, and is slower due to increased data handling and processing of each image. As a result, full spectral rasters are currently limited to a maximum of 30 exposures, corresponding to sizes of 3\arcmin\ or less (depending on the slit size).

For short exposure times or particularly high-cadence measurements, fewer spectral profiles per spatial pixel may be used to save telemetry. The amount of on-board data compression applied can also be adjusted, to obtain a balance between data quality and data volume.  For example, using the above 'dynamics' study: (a) 8 line profiles, 4:1 compression, 4.02~MiB volume per repeat, 181~kbps rate; (b) 4 line profiles, 9.84:1 compression, 0.82~MiB volume per repeat, 37~kbps rate.

The best compression settings to use are currently being studied in-flight.  Overall, SPICE provides a large amount of flexibility for designing science observations, and can be optimised to suit  different types of  measurements.

\subsubsection{Very bright sources in the quiet Sun network}

\begin{figure*}
\centering
\includegraphics[width=1.0\hsize]{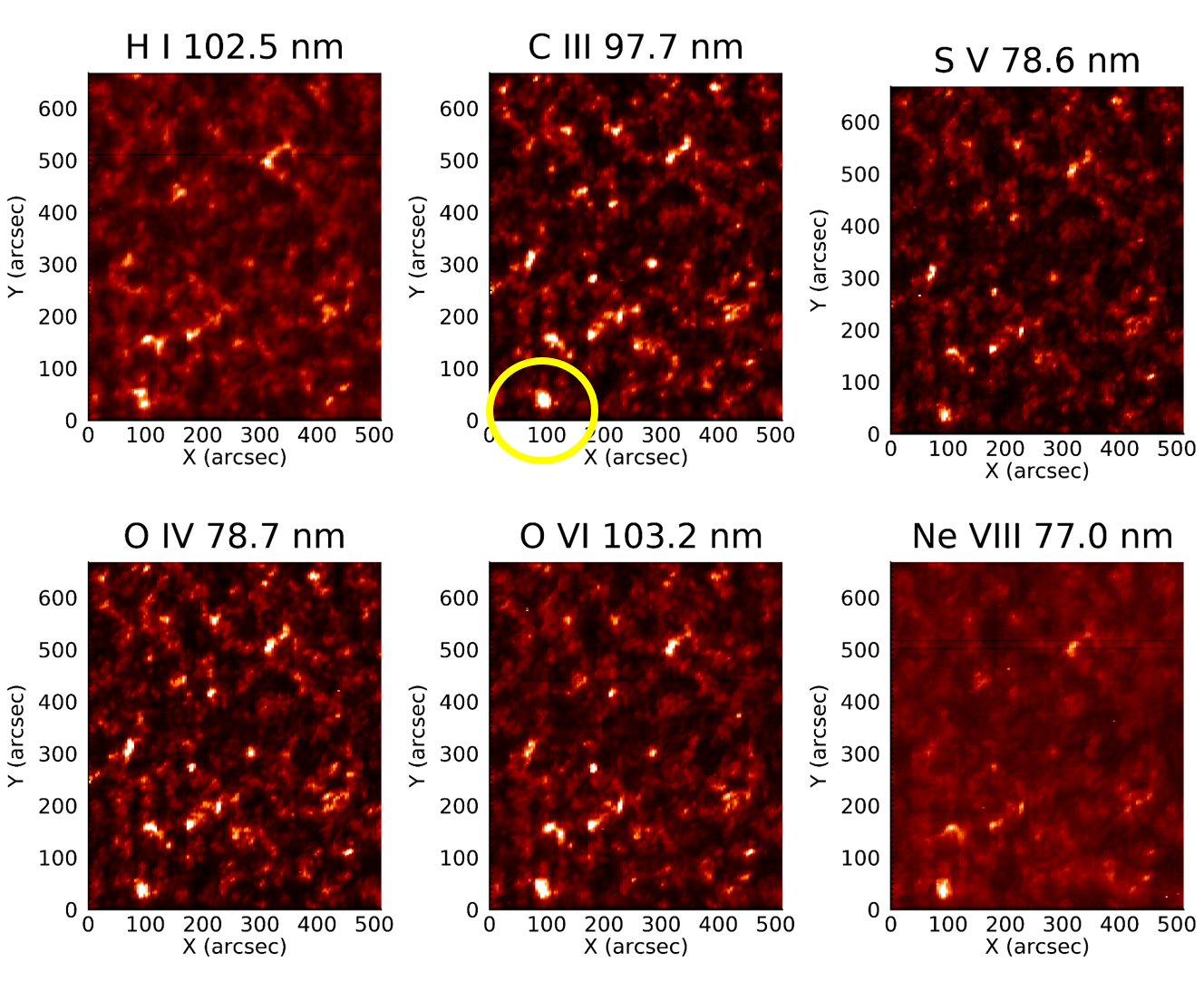}
\caption{Example of rastered images in selected lines from a quiet Sun area on 28 May 2020 from 16:05:37 to 16:50:24, using a 4\arcsec\ slit and 20~s exposure. Solar distance: 0.57~AU, 1{\arcsec}=413~km on the surface.  Images are arranged in the order of increasing formation temperature, from top-left: H~Ly~beta -- 10,000~K, C~III -- 60,000~K, S~V -- 140,000~K, O~IV -- 140,000~K, O~VI -- 260,000~K, Ne~VIII -- 550,000~K. The yellow circle shows the location of the brightest source (a 'beacon' – see text for details).}
\label{fig:3}
\end{figure*}

\begin{figure*}
\centering
\includegraphics[width=1.0\hsize]{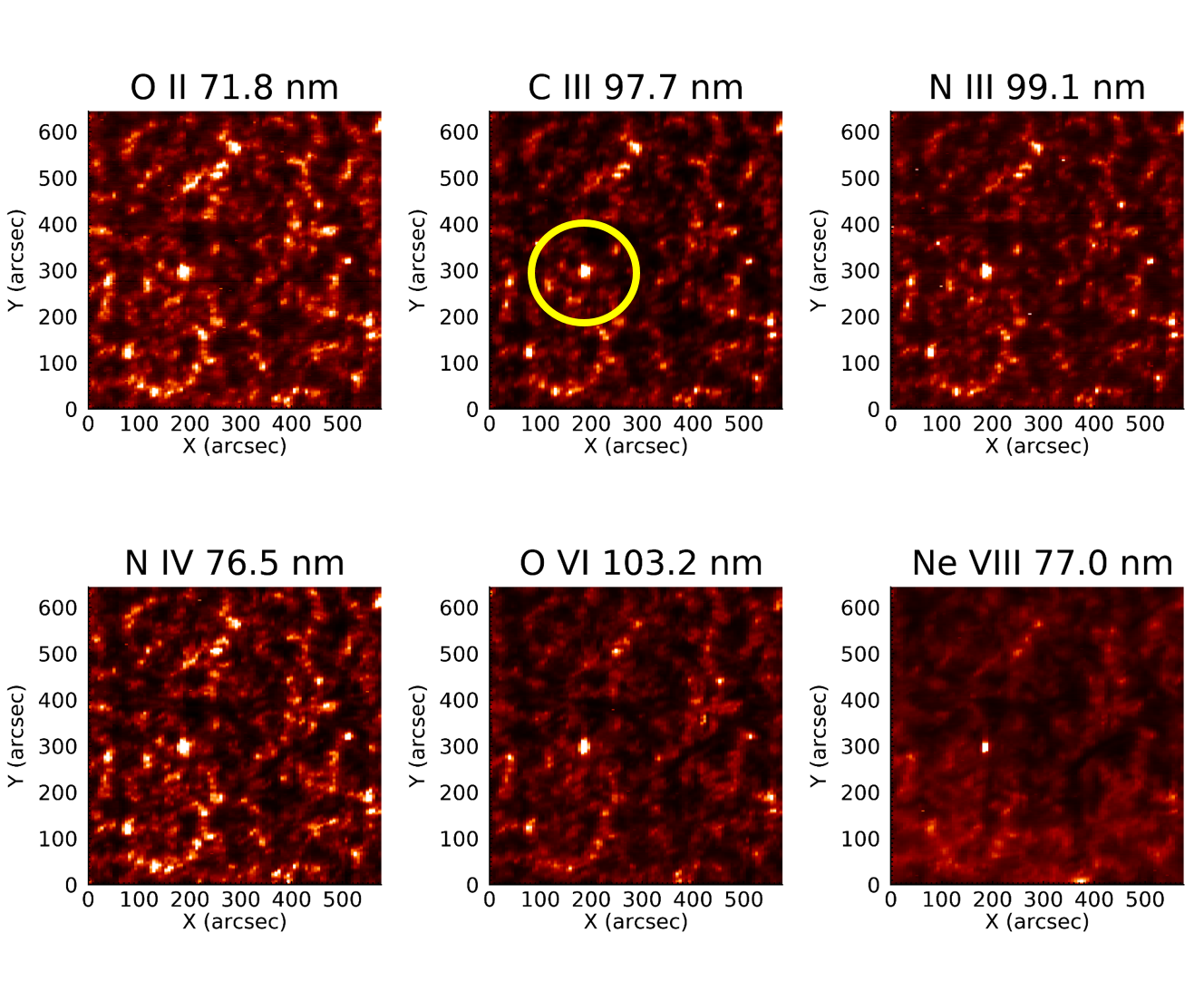}
\caption{Rastered images of the transition network on 17 June 2020 from 09:06:38 to 10:44:13~UT (solar distance: 0.52AU, 1\arcsec = 377~km on the surface) in a different set of lines, arranged in increasing order of temperature: O~II -- 40,000~K, C~III -- 50,000~K, N~III -- 66,000~K, N~IV -- 120,000~K, O~VI -- 260,000~K, Ne~VIII -- 550,000~K. The yellow circle shows the location of the brightest source -- see text for details. The observation was made using a 6\arcsec\ slit and 60~s exposure.}
\label{fig:4}
\end{figure*}

Figure \ref{fig:3} shows a selection of rasters in 6 bright lines taken at the Sun centre, using a 4\arcsec\ slit and 20~s exposure. Most of the low transition region images in the temperature range 50,000~K -- 300,000~K look similar, showing the typical chromospheric network extending into the transition region, with concentrations of brightness near network boundaries, as it has been abundantly observed before. Two lines that deviate most are the H~I~Ly$\beta$, due to a complex process of its formation at low temperatures, and the Ne~VIII line at 550,000~K because at these higher temperatures and greater heights the magnetic field starts diverging and the structures become fuzzier.

It is known from observations by the SUMER and CDS spectrometers on SOHO that the intensities of the transition region lines across the image area have a lognormal distribution \citep[e.g.][]{Pauluhn:2000}. It is therefore expected to see a number of quite bright concentrations of intensity in the network boundaries coming from the tail of the lognormal distribution of transition region intensities. For example, N~V 123.8~nm intensities from SOHO/SUMER plotted in \citeauthor{Pauluhn:2000}~(\citeyear[their Figure~4]{Pauluhn:2000}) extend to values approximately up to 10 times greater than the peak of their distribution. Intensities of the Ne~VIII 77.0 nm line (T$_e$ = 550,000 K) show enhancement up to 6 times greater than the peak of the distribution, and He~I 58.4 nm line up to a factor of 7. Similarly, \cite{Kretzschmar:2004} shows a lognormal distribution of N~IV 76.5~nm line intensity with an enhancement up to a factor of 10 above the peak of the distribution, and the H~I~Ly 91.8~nm line with a factor of 7 enhancement.

However, in some SPICE datasets we have found unusually bright sources with even more extreme intensity, by up to 25 times brighter than the average quiet Sun intensity and over 30 times greater than the peak of the intensity distribution. In this paper we initially refer to these sources as 'beacons'. We use this term to denote "a compact structure with the most extreme intensity that stands out in the observed quiet Sun area", without implying that these structures are different in nature from other transition region features known previously. It is just an indication of their extreme brightness. A discussion of similarities to blinkers, bright points and microflares is given below.

One of these beacons is seen in the bottom left corner of each image in Figure \ref{fig:3} and shown inside the yellow circle in the bottom-middle panel. Another example of a bright source is seen in Figure~\ref{fig:4}, near the centre of the images. It is present in all six images, particularly conspicuous in the Ne~VIII image. The beacon’s location is marked with a yellow circle on the C~III image.

If such sources were seen only in one raster, we might wonder whether they are short-lived, microflare-type events. However, on 14 May 2020 a beacon was observed in three consecutive rasters, spanning a time interval of 2~h 48~min. Another two beacons, on 28 May 2020 (Figure~\ref{fig:3}) and 17 June 2020 (Figure~\ref{fig:4}), were seen in two rasters, 45 minutes and 78 minutes apart, respectively, although the latter source on 17 June substantially decayed in brightness between the two observations. Therefore, these three bright sources are long-lived, slowly evolving features which are part of the network structure. They appear rarely and, in statistical terms, they are extreme outliers of the lognormal distribution of intensities (see Figure~7a). The beacons are therefore of interest to understand the maximum brightness that can be generated in the quiet Sun network. The ratio of the beacon’s peak intensity to the average intensity across the raster area varies with temperature -- it is largest between 50,000 and 300,000~K, where we get ratios as high as 18 to 25. The ratios become smaller, down to about 10, both below (He~I~Ly$\beta$) and above this temperature range (Ne~VIII). Table~\ref{tab:2} shows the ratio of maximum beacon intensity to the average, median, and the peak of intensity distribution across the SPICE raster in the N~III 99.1~nm line. This line, emitted at 70,000~K, has some of the largest ratios among the lines we use. We list all three measurements taken on 14 May 2020, showing the evolution of the bright source reaching the maximum brightness in the second (out of three) dataset:  with the maximum/median value reaching 22, while maximum/peak value reaching 32. For comparison, we show the same ratios for the 17 June 2020 data from Figure~4 where we also recorded the N~III line. There, the maximum/average ratio is 20, maximum/median is 25, maximum/peak is 38. 

 Table~3 compares these ratios for two dates, 14 May and 28 May 2020, for a strong line of C~III 97.7~nm, emitted at similar temperatures as the N~III line, around 65,000~K. This line achieves the largest ratios of 18 (max/median) and 36 (max/peak) in the 28 May 2020 data from Figure~\ref{fig:3}. Figure~\ref{fig:7}a illustrates how the maximum brightness (on the far right of Figure~7a) compares to the main intensity distribution across the raster and is further discussed in Section~2.3.2.

\begin{figure}
\centering
\includegraphics[width=1.0\hsize]{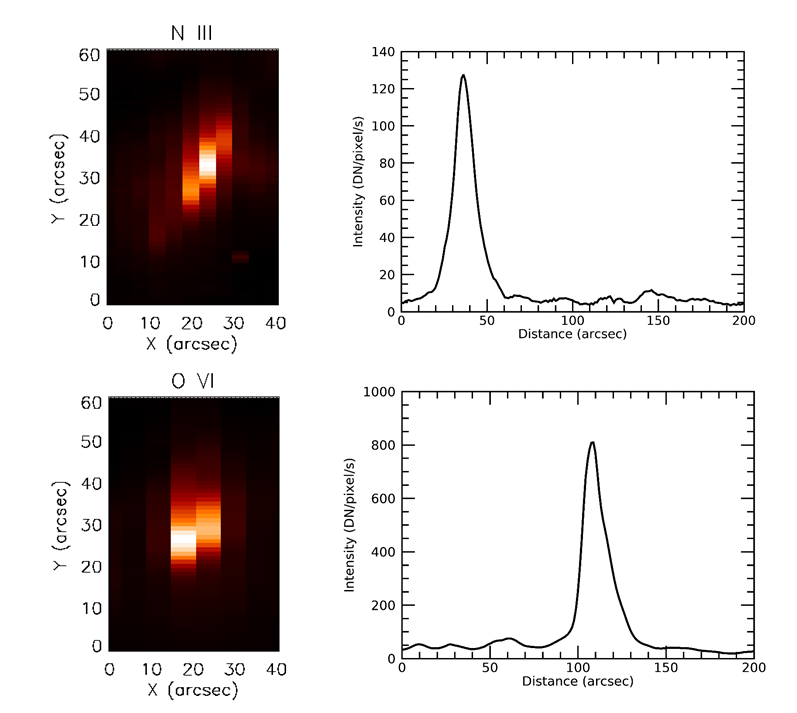}
\caption{(a) Left panels: close-up images of two beacons in the N~III line on 14 May 2020 (top) and O~VI line on 17 June 2020 (bottom), (b) Right panels:  distribution of intensities of the two beacons along a portion of the slit, for the X-position in the images corresponding to the brightest emission. Note that the top and bottom panels represent two different sources.}
\label{fig:5}
\end{figure}

Zooming in on two examples of beacons (Figure \ref{fig:5}), we can see they occupy an area only up to two or three slit positions wide, that is up to 12\arcsec. One of them is substantially concentrated inside the 4\arcsec\ slit. Their intensity distributions along the slit, shown on the right of Figure \ref{fig:5}, have a full width at half maximum of about 15\arcsec\ (6200~km and 5600~km on 14 May and 17 June 2020, respectively).  Given the SPICE instrumental point spread function (FWHM $\approx$ 6.7\arcsec), the actual sources have FWHM of ~12\arcsec\ along the slit. 

\begin{table*}[]
    \centering
    \caption{Relative brightness of the bright source in the N~III 99.1~nm line, in three observations made on 14 May 2020 and one observation on 17 June 2020. }
        \begin{threeparttable}[t]
    \begin{tabular}{|c|c|r|r|r|}
        \hline
        Date & Time (UTC) & Max/Ave & Max/Med & Max/Peak\\
        \hline
        14 May 2020  & 11:07:29 & 11.2 & 13.4 & 18.0 \\
                     & 11:57:37 & 18.4 & 22.7 & 31.9 \\
                     & 13:55:32 &  9.3 & 11.4 & 10.8 \\
        17 June 2020 & 09:06:38 & 20.0 & 24.8 & 38.0 \\
        \hline
    \end{tabular}
    \begin{tablenotes}
   \item[] The second column gives the time when the slit was positioned on the beacon. The next three columns show the ratio of the maximum beacon intensity (i.e. in the brightest pixel) to: the average intensity across the image area (Max/Ave); the median intensity of the image area (Max/Med), and the intensity corresponding to the peak of the intensity histogram (Max/Peak).
\end{tablenotes}
\end{threeparttable}
    \label{tab:2}
\end{table*}

\begin{table*}[]
    \centering
    \caption{ Comparison of the intensity enhancement of the bright source in the C~III 97.7~nm line, in observations made on 14 May 2020 and 28 May 2020. Description of columns is the same as in Table~2.  }
    \begin{tabular}{|c|c|r|r|r|}
        \hline
        Date & Time (UTC) & Max/Ave & Max/Med & Max/Peak\\
        \hline
        14 May 2020  &  12:13:55 & 12.6 & 15.6 & 28.1 \\
        28 May 2020  &  16:05:00 & 14.3 & 18.6 & 36.2  \\  
        \hline
    \end{tabular}
    \label{tab:3}
\end{table*}

\subsubsection{Relation of very bright sources to other observations} 

Below we consider similarities of beacons to previously described phenomena such as blinkers, coronal bright points and small explosive events. Transition region and coronal brightenings have been investigated in great detail before, and the question arises whether the strong brightenings seen here in SPICE are just a variant of a known feature or a new phenomenon. The brightenings we observe here are stretching over a large range of temperatures, essentially seen from the chromospheric to upper transition region emission simultaneously (cf. Figure~\ref{fig:3}) and have lifetimes of over one or two hours (because they are visible in successive raster maps). Thus, these brightenings are most likely not related to explosive events \citep[e.g.][]{Dere:1989,Innes:1997} because those are transients lasting only about a minute and do not seem to have an obvious coronal counterpart. More recently UV bursts have been found in active regions and a few also in the quiet Sun \citep[e.g.][]{Peter:2014,Young:2018}, but again, these are too short and (mostly) without coronal counterparts.

Our bright sources are also very different from ‘campfires’ seen by the Solar Orbiter EUI instrument \citep{Berghmans:2021}. The two main differences are: (a) the lifetimes of bright sources are longer than one hour while campfires have durations between 10~s and 200~s, with the shortest durations occurring most frequently; (b) the area of bright sources is 50-75~Mm$^2$ (see Figure~8), while the largest campfires have an area of only 3~Mm$^2$.

Blinkers, first described by \cite{Harrison:1997} based on observations from the SOHO Coronal Diagnostic Spectrometer (CDS), are an interesting but still rather unlikely candidate. They are compact features that show transient intensity enhancements in EUV emission lines -- just like explosive events they are restricted to transition region temperatures, and are most prominent in lines formed around 0.1--0.3 MK. Lifetimes are typically 16 minutes but can be as long as 40 minutes, and the intensity enhancement factors are typically a factor two, but can be as large as 20 \citep{Bewsher:2002}. They are expected to be observed in the SPICE transition region lines, and we could ask whether some of the beacons might correspond to the brightest blinkers. However, blinkers can only be definitively identified in datasets with a cadence of 5 minutes or better so we cannot confirm this connection. The fact that multiple beacons are identified in consecutive rasters separated by tens of minutes strongly suggests that these are long-lived structures rather than transient brightenings.

\begin{figure*}
\centering
\includegraphics[width=1.0\hsize]{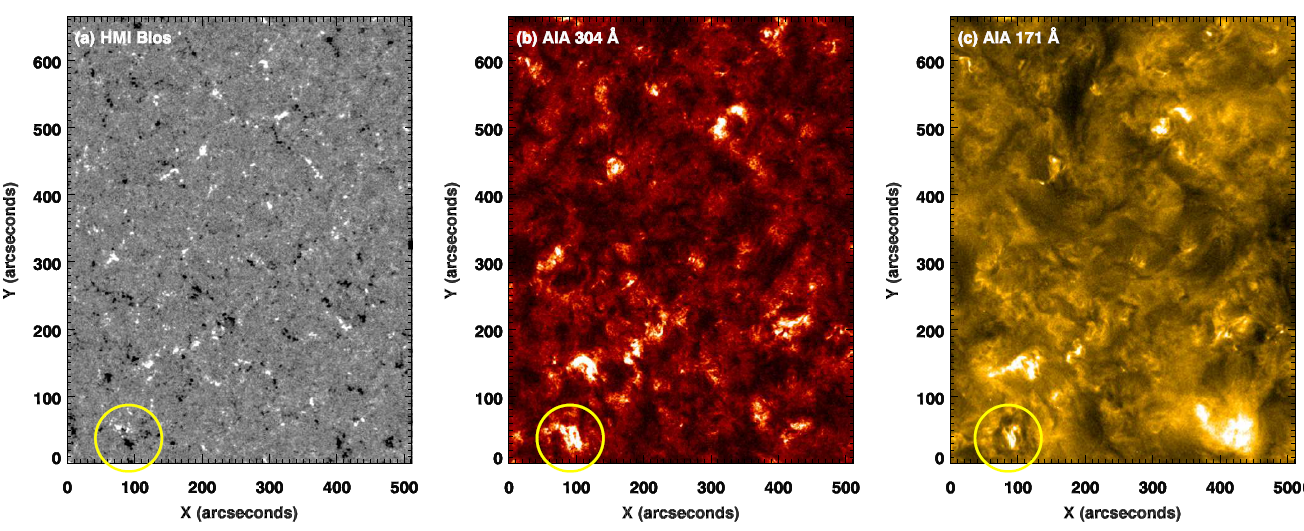}
\caption{Data from SDO on 28 May 2020 at 16:45 UT, of the same region as the SPICE raster shown in Figure~\ref{fig:3}. Here we show (a) the photospheric line-of-sight magnetogram from HMI, saturated at $\pm$50\,G, (b) the 30.4~nm channel from AIA corresponding to the transition region emission as seen by SPICE in C~III, and (c) the 17.1~nm channel from AIA corresponding to the coronal emission as seen by SPICE in Ne~VIII. The yellow circle highlights the brightening also highlighted in Figure~3. The SDO maps are projected onto the area as seen by SPICE and interpolated to a pixel scale of 435~km on the Sun that corresponds to AIA image scale of 0.6\arcsec\ per pixel.}
\label{fig:6}
\end{figure*}

A likely candidate to match the strong brightenings seen in the SPICE rasters are coronal bright points. A review by \cite{Madjarska:2019} characterises these to last several hours and to come in a wide range of sizes, also compatible with what SPICE sees here. Most importantly, coronal bright points show a simultaneous brightening over a wide range of temperatures, from the upper chromosphere to the hot corona \citep{Tian:2008}. In terms of relation to the magnetic field, coronal bright points occur above small bipolar structures, which eventually leads to an arcade of small coronal loops shaping the coronal bright point \citep{Madjarska:2019}.

For the data from SPICE showing the strong brightenings we do not have simultaneous observations of either the magnetic field or EUV emission from other instruments on Solar Orbiter (coordinated observations of all remote sensing observations can be expected in a regular fashion only with the start of the nominal mission phase in early 2022). The SPICE observations on 28 May 2020 were taken when Solar Orbiter was at an angle of about 30 degrees with respect to the Sun-Earth line, so luckily, we can use Earth-based facilities to obtain the surface magnetic field and transition region and coronal imaging. For this we use the Solar Dynamics Observatory \citep[SDO;][]{Pesnell:2012}, in particular the surface line-of-sight magnetograms from the Helioseismic and Magnetic Imager \citep[HMI;][]{Scherrer:2012} and EUV images from the Atmospheric Imaging Assembly \citep[AIA;][]{Lemen:2012}. For AIA we concentrate on the 30.4~nm channel dominated by He~II and the 17.1~nm channel dominated by Fe~IX. These correspond (in temperature) to the lower transition region line of C~III and the upper transition region line of Ne~VIII as recorded by SPICE. In Figure~6 we show the HMI magnetogram and the AIA images around 16:45 UT, the time when the SPICE slit crossed the brightening highlighted in Figure~\ref{fig:3}.

The strong brightening seen with SPICE in Figure~3 seems to match well with a typical bright-point structure. The magnetogram (Figure~\ref{fig:6}a; yellow circle) shows a small bipolar region with the main polarities separated by about 25\arcsec\ as seen from Solar Orbiter, which corresponds to a little more than 10~Mm on the Sun. Comparing the magnetogram to the H~Ly$\beta$ map (Figure~3), it is evident that the two small brightenings in H~Ly$\beta$ are co-spatial with the magnetic concentrations, as expected. The coronal image in AIA 17.1~nm (Figure~6c) shows an elongated structure seemingly connecting the two opposite polarities, indicative of a small loop arcade (even though the spatial resolution is not sufficient to clearly distinguish individual loop features). Interestingly, one of the two main polarities also contains a small patch of opposite polarity (small black spot in white region in yellow circle in Figure~\ref{fig:6}a). This might be indicative of small-scale reconnection powering the bright point in a fashion similar as reported by \cite{Chitta:2017}. Following the temporal evolution of this feature shows that it lasts for significantly longer than one hour. During this time the brightness changes, but always remains significantly above the average background emission. This provides circumstantial evidence that the strong brightening seen in SPICE on 28 May 2020 and, by analogy, the other two strong brightenings as well, indeed correspond to coronal bright points.

We point out, however, that while the coronal bright point in the AIA 171 image (Figure~\ref{fig:6}) is both smaller and fainter than several other coronal features in that image, its transition region counterpart in SPICE is the brightest source in the SPICE field of view.  

Having identified the bright source (the beacon) in the image  on 28 May as located under a coronal bright point, we can expand the discussion from Section 2.3.1 and compare its intensity to the distribution of intensities in all pixels in the C~III image for 28 May. Figure~7a shows that the majority of pixels have a lognormal distribution of intensities, illustrated by the red curve – this is typical of the quiet Sun as already pointed out in Section 2.3.1. While we are interested in the brightest source (i.e. the bright point), we note that there may be other, lower intensity sources with intermediate intensities, still above the lognormal curve. 
A small number of pixels have large intensities, significantly deviating from the lognormal curve, and are seen at the far right in Figure~7a – this includes the beacon and a few other sources.  Figure~7b shows locations of all C~III image pixels with intensities above 400.  There are 18 small sources above this threshold.  

Figure~8 shows the maximum intensities of these 18 sources in the C~III line above the 400 threshold for 28 May, sorted by decreasing intensity. It reveals that there are two very bright sources in the 28 May 2020 raster, numbered 1 and 2. In fact, the maximum intensity of source \#1 is slightly greater, but it has a significantly smaller area and lower total intensity than source \#2, our original beacon discussed earlier. Both sources are labelled in Figure~7b, as ‘1’ and ‘2’. Comparing Figure~7b to the HMI magnetic field in Figure~6, we find that source \#1 is also associated with a small bipolar magnetic field structure and could be classified as a small bright point. Therefore, using a source intensity as a criterion, it is sometimes possible to find more than one beacon (bright point) in the raster area at any given time. This should not be too surprising as it entirely depends on the distribution and evolution of the photospheric magnetic field, and our classification criterion of ‘extreme brightness’ is of a qualitative nature. To make the criterion more stringent, one could define an intensity enhancement level, for example, the ratio of the maximum intensity to the median intensity greater than a prescribed value.  Table~2 and~3 illustrate that a value above 15 would include the enhancement levels seen in the tables that reach up to 25, depending on the selected spectral line and on the particular data set. Further work based on a larger sample of events will yield a more precise definition of bright points in the transition region that can be generally applied to quiet Sun datasets.

The red curve in Figure~8 shows areas of these 18 sources, sorted in the same order as maximum intensities. The area is expressed as the number of pixels with a 1\arcsec $\times$ 1.1\arcsec\ size  (one pixel has an area of 0.188~Mm$^2$ at the distance to the Sun on 28 May 2020). The main beacon (\#2) has the largest area (75~Mm$^2$), while the remaining smaller sources gradually decrease from 28~Mm$^2$.  Comparing the areas of the brightest sources in the other two data sets, they are 58~Mm$^2$ and 54~Mm$^2$ on 14 May 2020 and 17 June 2020, respectively.

\begin{figure*}
\centering
\begin{subfigure}[h]{0.49\hsize}
         \centering
         \includegraphics[width=\textwidth]{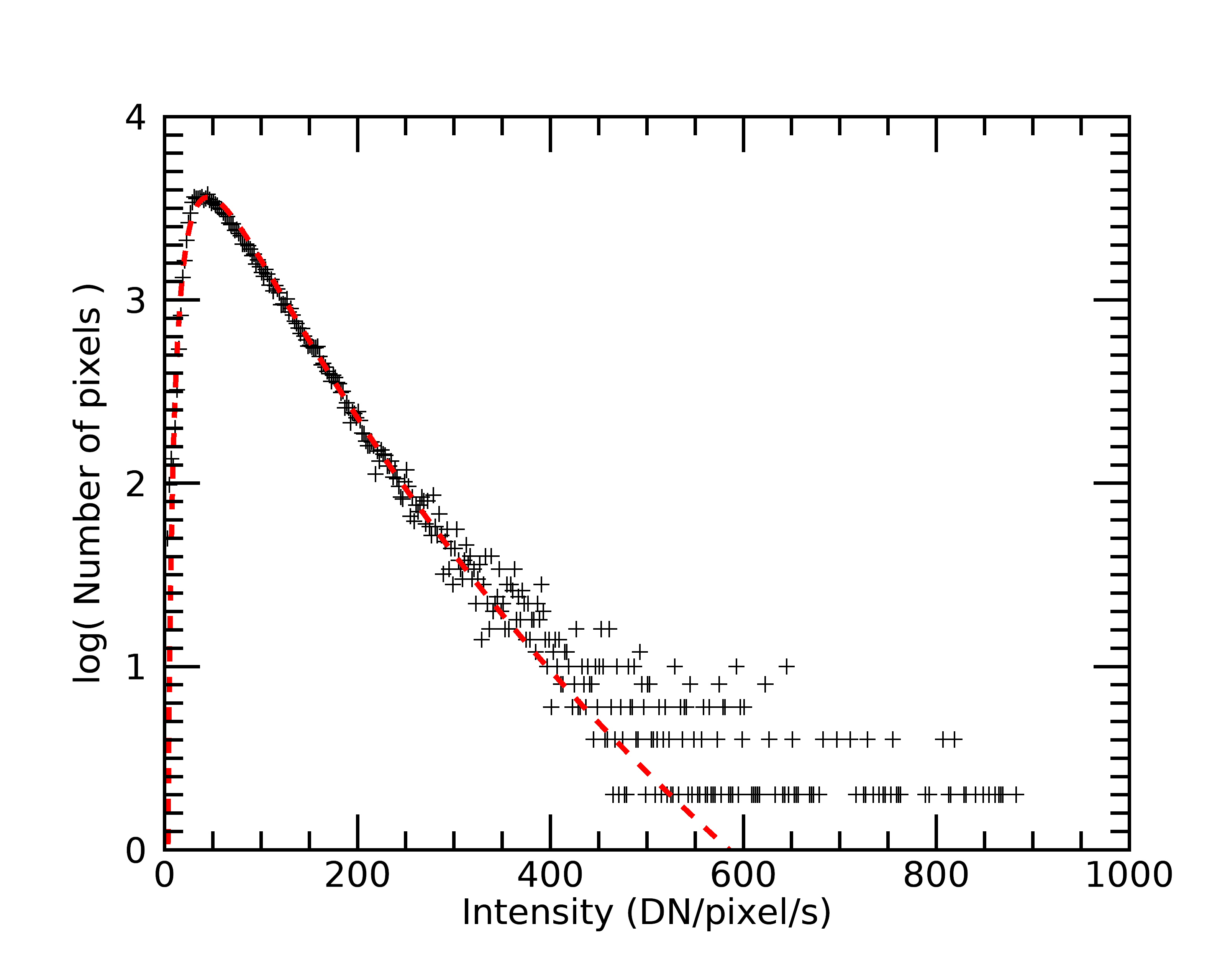}
     \end{subfigure}
      \hfill
     \begin{subfigure}[h]{0.49\hsize}
         \centering
         \includegraphics[width=\textwidth]{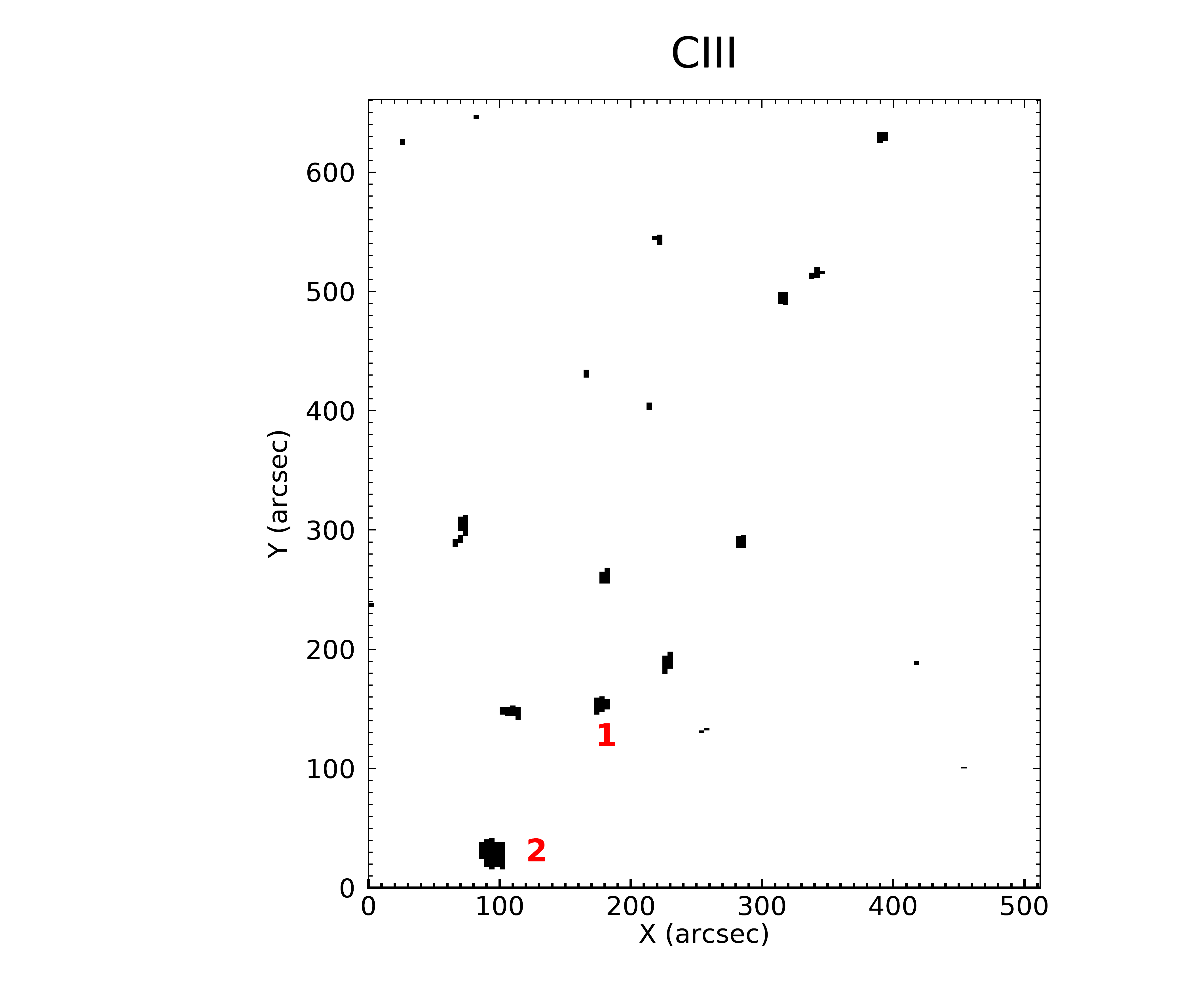}
     \end{subfigure}
\caption{Results for the SPICE raster from 28 May 2020 between 16:05:37 to 16:50 UT, in the C~III line shown in Figure~\ref{fig:3}. (a) Left panel: A distribution of intensities in C~III across the entire raster in 1\arcsec$\times$1.1\arcsec\ pixels (each 4\arcsec-wide pixel has been divided into four 1\arcsec-wide pixels). The red dashed line shows a lognormal fit to the main distribution. The points on the far-right deviating most from the red line include two bright points with the highest intensities, marked '1' and '2' in panel~(b); (b) Right panel: An image of the SPICE raster area showing the locations of all sources in the C~III line with intensities above a threshold of 400 from panel (a). The two bright points with the highest maximum intensity are marked. }
\label{fig:7}
\end{figure*}

\begin{figure}
\centering
\includegraphics[width=1.0\hsize]{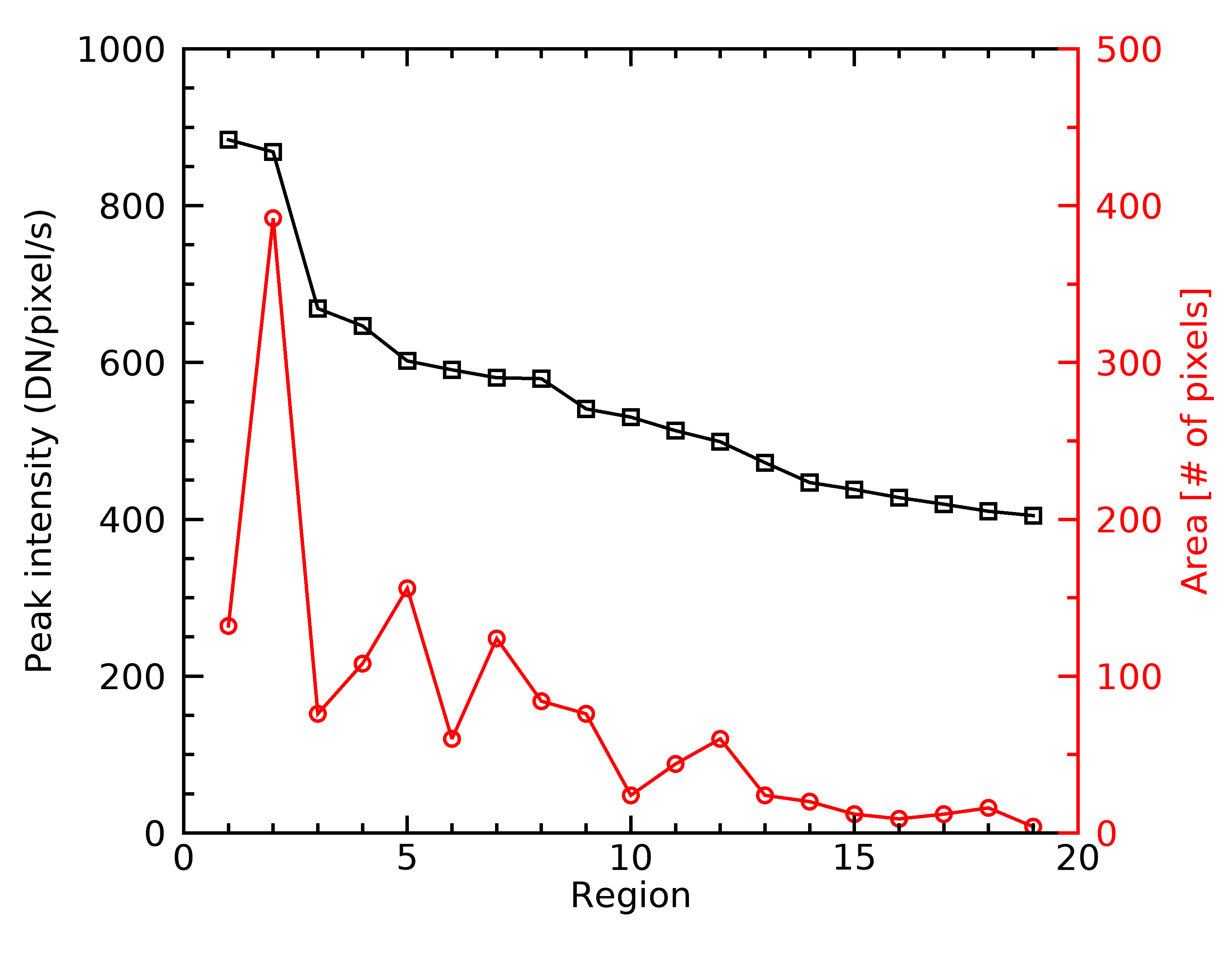}
\caption{Parameters of the sources from Figure~\ref{fig:7}b for the SPICE C~III line on 28 May 2020 at 16:05~UT. Black curve (squares): source peak intensity, sorted in a decreasing order;  Red curve (circles): source area, expressed as a number of 1\arcsec$\times$1.1\arcsec\ pixels, sorted in the same order as the peak intensity. One pixel corresponds to 0.188~Mm$^2$.}
\label{fig:8}
\end{figure}

Future studies with SPICE in combination with EUI and PHI on Solar Orbiter (and with Earth-based facilities) will provide interesting possibilities to study these features. SPICE can offer diagnostics continuously from the chromosphere to well above a million K and thus offers a broader temperature range than other single instruments. This will allow a more reliable emission measure analysis of the thermal structure of bright points. With high-resolution observations from EUI and PHI available on the same platform, such combined data should shed new light on the structure and evolution of coronal bright points. 

\subsubsection{Limb observations}
On two occasions, 28 May and 20 June 2020, the spacecraft was pointed to the limb, visiting all four cardinal points, north, south, east and west. These manoeuvres provided the first opportunity for SPICE to take measurements above the limb.

\begin{figure}
\centering
\includegraphics[width=1.0\hsize]{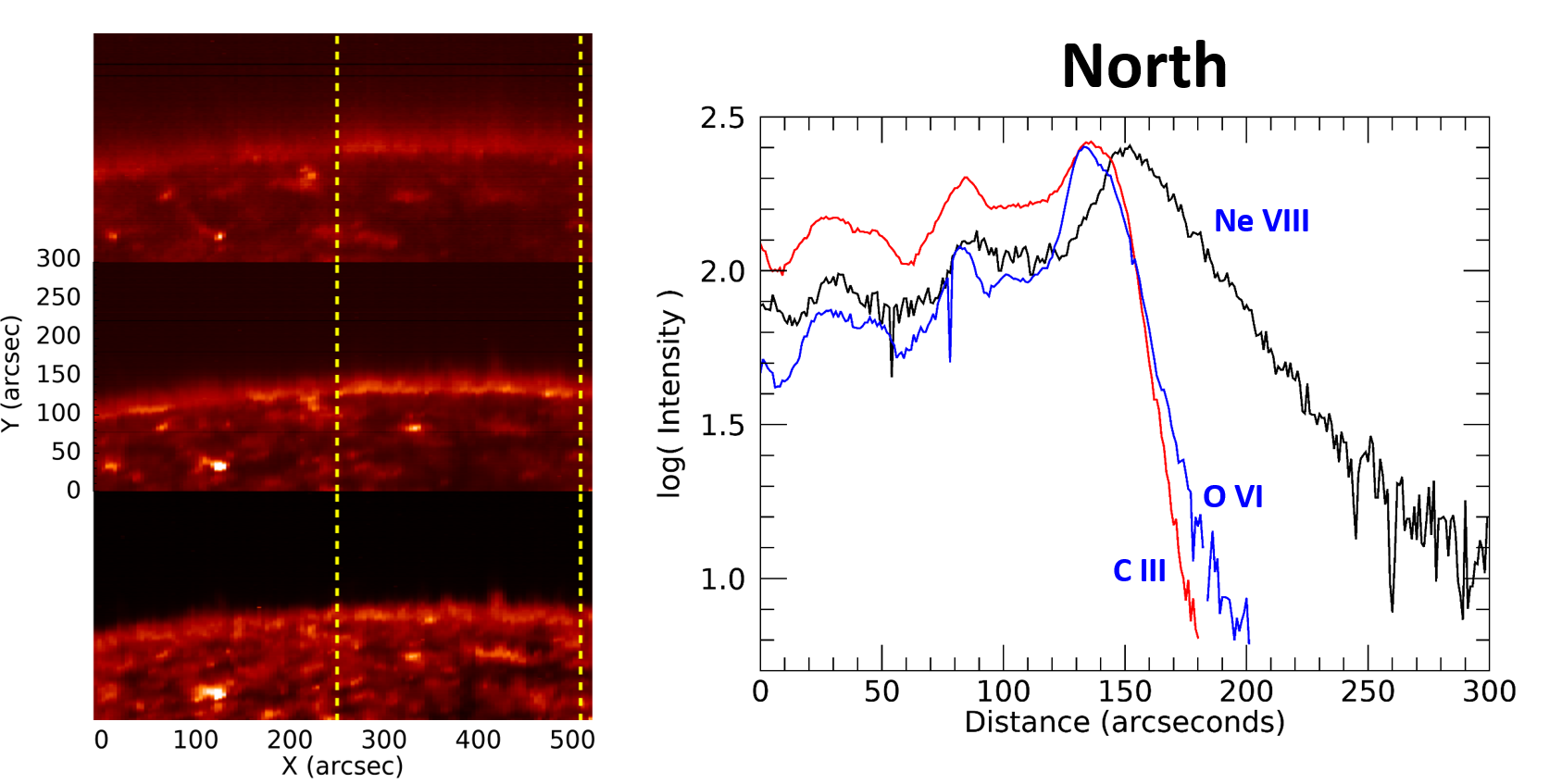}
\caption{Limb images at the north pole (solar distance: 0.57~AU, 1\arcsec = 413~km on the surface). (a) Left panel: from the bottom, C~III, O~VI, Ne~VIII. The yellow dashed lines show the E-W range where the line intensities in panel~(b) have been summed. (b) Right panel: intensities of the three lines with distance above the limb, averaged over 250 pixels in the E-W direction. The limb is at 130{\arcsec}.}
\label{fig:9}
\end{figure}

SPICE observations above the north polar region are shown in Figure~9 in C~III 97.70~nm, O~VI 103.2~nm and Ne~VIII 77.0~nm lines. Figure~9 also shows the intensity decrease with height above the north limb for the three lines. As expected, the coolest line of C~III drops very fast with height, while the hottest line of Ne~VIII decreases less steeply and can be seen at least 2 arcminutes (0.07~R$_{\odot}$) above the limb.  This will allow us to measure the coronal emission at these heights and estimate the electron density, assuming the Ne~VIII intensity is proportional to density squared. The Ne~VIII intensities above polar coronal holes can help with the modelling of the properties of the fast solar wind.

The Ne~VIII intensity at the north limb was compared to its peak intensity at the west equatorial limb, and found to be 2.9 times lower, suggesting that the coronal electron density at the polar limb was 1.7 times lower. Other, cooler lines showed only a 10--20\%\ difference between polar and equatorial limb intensities. 
\citeauthor{Stucki:2000}~(\citeyear[their Figure~6]{Stucki:2000}) find a similar behaviour of lines cooler and hotter than 550,000~K observed by SOHO/SUMER, in particular, the Ne~VIII line intensity in a coronal hole near the limb has intensity reduced by approximately a factor of 2.8 compared to a quiet Sun area on disk. 

Similar off limb observations \citep{Fludra-Del-Zanna:1999} were made in 1996 with the SOHO/CDS, in the quiet Sun corona at the minimum of the solar cycle. They showed that the emission in coronal lines of Mg~IX, Mg~X, and Si~XII extends far above the equatorial limb, and the electron density decreases exponentially, starting from $5\times 10^8$ cm$^{-3}$. The electron density near the limb above polar coronal holes was a factor of two lower. The behaviour of the SPICE Ne~VIII line intensity described earlier is consistent with this. 

SPICE will be able to reach 8\arcmin\ above the east and west limb at 0.5~AU distance and closer, when the spacecraft is pointed to the corresponding limb (the maximum allowed off-point). On 26 March 2021, in a special observation taken outside the commissioning period, we have tested the sensitivity of SPICE at the greatest available height at the time, 5\arcmin, above the equatorial east limb.  A full spectrum was acquired with a 10 minute exposure when the spacecraft was at 0.72 AU from the Sun. At this distance of 0.22 solar radii above the limb, we can still see five spectral lines:  Mg~IX 70.60~nm, two Ne~VIII lines at 77.0 and 78.0~nm, and two O~VI lines at 103.2 and 103.8~nm. In the lower panels of Figure~10 we show these profiles obtained by averaging 200 pixels along the length of the slit. We have tested that the brightest cool lines of H~Ly$\beta$ and C~III have no detectable signal (above the background noise) at this distance, or at any distance greater than 2.5\arcmin\ (0.09 solar radii) above the limb.  This compares to 171 and 212 DN/s on-disk (see Table~1) for H~Ly$\beta$ and C~III, respectively, thus demonstrating the absence of instrumental scattering from the disk in this off limb measurement.  The signal in the plotted lines is therefore a real emission from the corona.

The O~VI line pair off limb, in both polar and equatorial region, is particularly interesting because it was used in the past by coronagraphs, such as UVCS \citep{Noci:1987,Kohl:1997} to measure the outflow velocity of the solar wind using the Doppler dimming effect. This method requires a measurement of the intensity ratio of the two lines, combined with appropriate modelling.  There is limited data from previous missions at distances close to the limb \citep{Teriaca:2003}, and further investigations with SPICE will identify how well it can contribute and to what level of accuracy.  This is a subject of a separate study and requires more calibrated data.

\begin{figure}
\centering
\includegraphics[width=1.0\hsize]{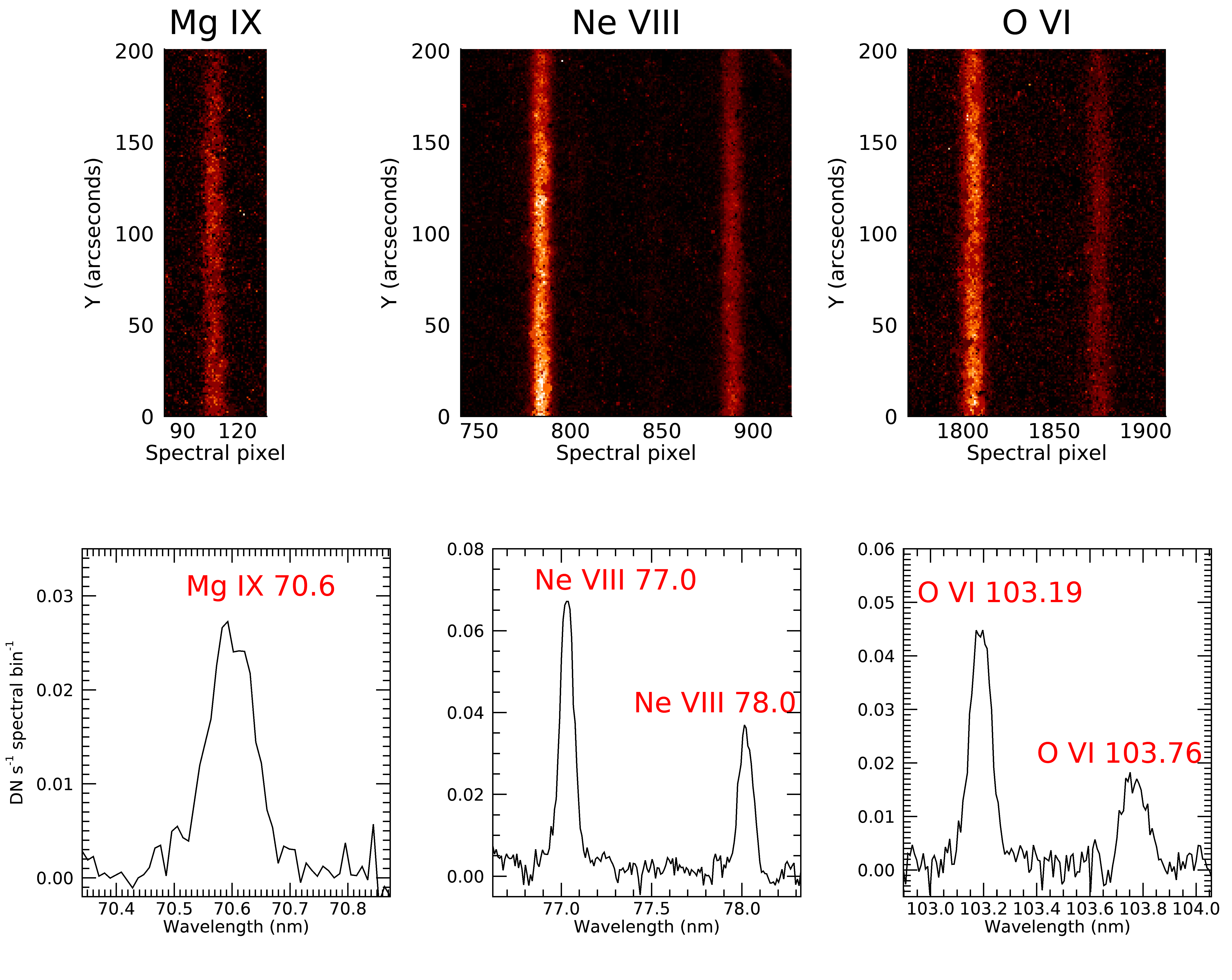}
\caption{Observations at 0.22 solar radii above the east limb on 26 March 2021, using 10 minute exposure.  Upper panels: image on the detector, along part of the slit, of Mg~IX 70.60~nm, Ne~VIII 77.0 and 78.0~nm, O~VI 103.2 and 103.8~nm. Lower panels: profiles of Mg~IX 70.60~nm, Ne~VIII 77.0~nm and 78.0~nm, and O~VI 103.2~nm and 103.8~nm, averaged over 200 pixels along the slit.}
\label{fig:10}
\end{figure}

\subsubsection{Elemental abundance diagnostics}
It is long-standing knowledge that the coronal abundances of some elements are different from their photospheric abundances \citep{Meyer:1985}. Both spectroscopic and solar energetic particle data show that the coronal-to-photospheric abundance ratios of elements with low first ionisation potential (FIP $<$ 10~eV) are enhanced by up to a factor of four relative to those with high FIP \citep[e.g.][]{Fludra-Schmelz:1999, Schmelz:2012}. This fractionation is known as the 'FIP effect'.  It is now believed that the low-FIP elements increase their absolute abundances (i.e. relative to hydrogen) in the corona \citep{Feldman:2002}, while the absolute abundances of high-FIP elements in the corona remain the same as their photospheric values. The coronal composition can also vary from region to region, or in time. 

These composition variations will be a key measurement from SPICE and can be quantified by measuring the 'FIP bias', defined as the ratio of an element’s abundance in the upper solar atmosphere to that element’s abundance in the photosphere. It is expected that the FIP bias maps obtained by SPICE in the transition region and the low corona can be correlated with in~situ measurements of heavy ions made by the Solar Wind Analyser/Heavy Ion Sensor \citep{Owen:2020} on Solar Orbiter. This will help to confirm the magnetic connectivity between the spacecraft location and the observed area on the Sun’s surface. Spectroscopic observations in the solar atmosphere and in~situ data can be then combined \citep{Fludra:2018} to test solar wind model predictions, in order to assess whether outflows measured in the corona originate from open field line structures and extend into the solar wind.

SPICE is the only Solar Orbiter instrument capable of deriving the FIP bias in the solar upper atmosphere. Different diagnostics techniques exist to derive the FIP bias from measurements and are under investigation for SPICE. FIP biases are usually calculated using the line ratio of two spectral lines with similar formation temperature or through a differential emission measure analysis. Recently, \cite{Zambrana:2019} developed a new method to determine relative elemental abundances which relies on linear combinations of spectral lines optimised for FIP bias determination.

Figures \ref{fig:1} and \ref{fig:2} show that the SPICE wavelength bands are dominated by lines from high-FIP elements (O, C, Ne, N), but they also include a few lines from low-FIP elements, in particular Magnesium and Sulphur.  The latter is considered an intermediate FIP element (FIP=10.4 eV) but has been used to derive composition maps \citep[e.g.][]{Brooks:2015}.

The most promising diagnostics for measuring the low-FIP/high-FIP abundance enhancement and establishing the solar wind connection is the Mg/Ne ratio, which can potentially be obtained using lines arising from Ne~VIII, Mg~VIII and Mg~IX. This will provide information at temperature around 0.6-1.0 MK. At lower temperatures of 0.1-0.2 MK, the FIP effect can be investigated with Sulphur relative to Oxygen and/or Nitrogen using lines arising from S~III, S~IV, O~III to V and N~III, N~IV. The abundance analysis is a work in progress that requires a careful analysis of the temperature distribution. 

\section{Discussion and conclusions}
\label{section:discussion}
SPICE is the only instrument on board Solar Orbiter that can measure EUV spectra from the disk of the Sun and the low corona, and can record all spectral lines simultaneously at rates as fast as every second. This helps to monitor rapidly evolving solar features.

The SPICE data obtained between April and June 2020 have given us exciting observations of the first full spectra with 23 bright lines, images of the quiet sun transition region at sun centre and at the polar and equatorial limb at a wide range of temperatures, and a detection of beacons -- rare, compact bright sources with extreme intensities far above the typical lognormal intensity distribution. So far, we have found compelling evidence that these bright sources are a transition region signature of the coronal bright points. The detected bright sources can be used to place a limit on the intensity increase of transition region lines that is possible in the quiet Sun network and will stimulate a study of their relationship with the local magnetic field. 

The far off limb observations showed that SPICE can see five lines from three ions (Mg~IX, O~VI and Ne~VIII) at significant heights in the corona up to 0.22~R$_{\odot}$ above the equatorial limb. Therefore, potentially we can derive the relative Mg to Ne abundance once the absolute radiometric calibration of SPICE becomes available. When compared to the in situ ion composition measured by the SWA/HIS sensor, this will help to verify the magnetic connection between the Sun’s surface and the spacecraft.

The intensity of the Ne~VIII 77.0~nm line above the equatorial region (closed magnetic fields) is 2.9 times greater than above the polar regions with open magnetic field. This suggests that the electron density above the polar coronal holes is approximately 1.7 times lower, assuming that the line intensity is proportional to electron density squared.   

SPICE observations of the equatorial quiet Sun and polar regions between April and June 2020 were taken at the beginning of the solar cycle.  They can therefore be compared to SOHO observations at other solar minima, in 1996--1997 and 2007--2008, particularly those taken by the SOHO/CDS of the off-limb corona \citep{Fludra-Del-Zanna:1999}.  There, the electron densities of 1~MK coronal plasma in the polar region were a factor of two lower than the equatorial densities. The off-limb coronal hole/quiet Sun intensity ratios of the Ne VIII line are also close to those measured on disk by \cite{Stucki:2000}. 

One of the main goals of SPICE is to study the origin and sources of the fast solar wind inside polar coronal holes.  Therefore, one of our future targets is to observe polar coronal holes from higher latitudes, when the spacecraft climbs out of the ecliptic up to 30 degrees in the later part of the Solar Orbiter mission, in years 2025 -- 2029. This will allow SPICE to derive velocity maps and measure the outflows in polar coronal holes with improved spatial resolution. The study of Doppler velocities, as mentioned in Appendix \ref{appendix:performance}, will be a subject of another paper.  

In summary, the first observations presented in this paper have illustrated that SPICE can be used to study the following:
\begin{enumerate}
    \item The EUV spectrum in the range 70.4–-79.0 nm and 97.3--104.9 nm.
    \item Temperature distribution of transition region and low corona between 30,000 K and 1,000,000 K.
    \item Monitor the distribution of spatial structures in the chromosphere and transition region.
    \item Monitor the temporal variability of the chromospheric and transition region line intensities with cadence as fast as 5~s for sit and stare studies, or using rasters of arbitrary width up to 16 arcminutes, with exposure times of 5~s to 180~s. Usefulness of exposures as short as 1~s will be demonstrated in the future on bright targets such as active regions.
    \item Carry out further analysis to derive abundance ratios of low-FIP/high-FIP elements using Mg~IX, Mg~VIII and Ne~VIII lines, and also Sulphur lines compared to Oxygen and Nitrogen lines.
    \item To study unusually bright compact structures with extreme intensities in transition region lines (identified as coronal bright points), including their temporal evolution.
    \item To study bright lines above the polar limb, and with additional modelling, to measure the distribution of electron density with height.
    \item To make measurements of Mg~IX, Ne~VIII and O~VI lines at distances up to 0.22~R$_{\odot}$ from the equatorial limb. Further studies of above the limb measurements will show to what extent they can be used to constrain the properties of the solar wind.
\end{enumerate}

This list shows the richness of the type of observations that will be carried out by SPICE in the nominal mission phase, contributing to the Solar Orbiter mission goals.

\begin{acknowledgements}
Solar Orbiter is a space mission of international collaboration between ESA and NASA, operated by ESA. The development of the SPICE instrument was funded by ESA and ESA member states (France, Germany, Norway, Switzerland, United Kingdom). The SPICE hardware consortium was led by Science and Technology Facilities Council (STFC) RAL Space and included Institut d’Astrophysique Spatiale (IAS), Max-Planck-Institut f\"ur Sonnensystemforschung (MPS), Physikalisch-Meteorologisches Observatorium Davos and World Radiation Center (PMOD/WRC), Institute of Theoretical Astrophysics (University of Oslo), NASA Goddard Space Flight Center (GSFC) and Southwest Research Institute (SwRI). The in-flight commissioning of SPICE was led by the instrument team at UKRI/STFC RAL Space. AF’s and AG’s research is funded by UKRI STFC. ADAS is a project managed at the University of Strathclyde (UK) and funded through membership of universities and astrophysics and fusion laboratories in Europe and worldwide. The authors thank the referee for a careful reading of the manuscript and detailed comments that helped to improve the paper.
\end{acknowledgements}

\bibliographystyle{aa} 
\bibliography{First_Obs_SPICE.bib} 

\appendix
\section{Data processing}
\label{appendix:dp}
 
\begin{table*}[]
    \centering
    \caption{SPICE datasets discussed in this paper.}
    \begin{tabular}{|l|c|c|r|c|c|c|}
         \hline
         \textbf{Observation} & \textbf{Date/Start Time (UTC)} & \textbf{SPIOBSID} & \textbf{Exp. Time (s)} & \textbf{Slit} & \textbf{FOV} & \textbf{Windows} \\
         \hline
         Fixed, Sun Centre & {21-Apr-20 / 13:59} & 12583408 & 180 & 2\arcsec x 660\arcsec & 2\arcsec x 660\arcsec & Full spectrum \\
         Raster, Sun Centre & 14-May-20 / 09:39 & 12583472 & 60 & 6\arcsec x 660\arcsec & 576\arcsec x 660\arcsec & 6 \\
         Raster, Sun Centre & 14-May-20 / 11:41 & 12583474 & 60 & 4\arcsec x 660\arcsec & 256\arcsec x 660\arcsec & 6 \\
        Raster, Sun Centre & 14-May-20 / 13:19 & 12583475 & 120 & 2\arcsec x 660\arcsec & 256\arcsec x 660\arcsec & 6 \\
         Raster, Sun Centre & 28-May-20 / 16:05 & 12583687 & 20 & 4\arcsec x 660\arcsec	& 512\arcsec x 660\arcsec & 7 \\
         Raster, North Limb & 28-May-20 / 17:50 & 12583688 & 20 & 4\arcsec x 660\arcsec & 512\arcsec x 660\arcsec & 7 \\
         Raster, Sun Centre & 17-Jun-20 / 09:06 & 16777417 & 60 & 6\arcsec x 660\arcsec & 576\arcsec x 660\arcsec & 6 \\
         Fixed, East Limb & 26-Mar-21 / 23:33 & 50331900 & 600 & 4\arcsec x 660\arcsec & 4\arcsec x 660\arcsec &	Full spectrum \\
         \hline
    \end{tabular}
    \label{tab:A.1}
\end{table*}

The SPICE datasets used for the results presented in this paper are listed in Table~{A.1}. SPICE observations are uniquely identified by the SPICE Observation ID (contained in FITS keyword SPIOBSID),  which is used in the filenames of the corresponding data products.  In the case of a repeated series of rasters, the Raster ID number (contained in FITS keyword RASTERNO) is used to distinguish each repeat within the observation series. The flight data products, accessible via the Solar Orbiter archive, are available in two processing levels: (a) level~1: uncalibrated data in engineering units; (b) level~2: calibrated data in physical units.

Further details of the products, and the level 1 to level 2 processing steps are available in \cite{Spice-all:2020}. Note that the FITS file headers also contain WCS (world coordinate system) parameters to specify the observation coordinates in common solar units.  

The analysis conducted for this paper began with level 1 products, as the level 2 pipeline was still under development at the time.  The processing steps used are outlined below:
\begin{itemize}
    \item[1.] Dark current subtraction:
    \begin{itemize}
        \item In-flight dark data were used for this step, subtracting a dark image with an identical exposure time.  The nearest available dark image (in time) to the observation was used in each case. 
    \end{itemize}
    \item[2.] Pixel response (also known as `detector flat-field') correction:
    \begin{itemize}
        \item The detector flat-field was measured on ground \citep{Spice-all:2020}, and is still applicable in-flight.  This removes the variation in response for each pixel across the detector arrays.  The correction is applied by dividing each pixel by a number (always close to 1.0) from the flat-field calibration file.
        \item This correction does not include any detector burn-in (reduction in response due to prolonged solar exposure), as this is not needed for the first observations.  This correction will be determined and made available via the flight pipeline as it becomes needed.
    \end{itemize}
    \item[3.] Geometric distortion correction:
    \begin{itemize}
        \item A prototype correction algorithm has been developed to correct for the optical distortion and slant/tilt of the spectral and spatial axes relative to the detector pixels \citep{Thompson:2020}. 
        \item The data reduction used the same algorithm that is being integrated into the flight pipeline.
    \end{itemize}
    \item[4.] Wavelength calibration:
    \begin{itemize}
        \item A prototype version of the flight calibration was used to assign the wavelength to each detector pixel.
        \item As described in \cite{Spice-all:2020}, an initial version was determined from ground testing, which was found to vary slightly with instrument temperature. The in-flight variation has been measured at several solar distances (and hence temperatures) so far, and this has been applied to the data presented in this paper. Full details of the flight calibration will be the subject of a future paper.
        \end{itemize}
    \item[5.] Absolute calibration: not used for this paper. All data remained in uncalibrated units (detector counts). 
 
\end{itemize}
Additional steps (e.g. averaging, line fitting etc.) were carried-out for analyses of the specific observations, and these are described in the relevant sections of this paper. 

\section{Preliminary instrument in-flight performance}
\label{appendix:performance}
After characterising the focus adjustment of the telescope during commissioning, initial assessments were made of the instrument spatial and spectral resolution.  They can be compared to initial ground measurements, and flight performance predictions where measurements were not possible, reported in \cite{Spice-all:2020}: 
\renewcommand{\labelenumi}{(\alph{enumi})} 
\begin{enumerate}
    \item Spatial resolution: prediction -- 5.4{\arcsec}, flight -- 6.7{\arcsec};
    \item Spectral resolution (2\arcsec\ slit), SW channel: ground test -- 4.7 pixels, flight -- 7.8 pixels;
    \item Spectral resolution (2\arcsec\ slit), LW channel: ground test -- 5.3 pixels, flight -- 9.4 pixels.
\end{enumerate}

The geometric limitations of the optical ground test set-up were described in \cite{Spice-all:2020}, including a small test beam size compared to the instrument aperture. This may have contributed to the differences in measured ground and flight performance, but the reasons are still under investigation.
 
Possible root causes of the larger spatial and spectral widths in-flight are the focus and aberration effects: (1) a focusing error of the spectrometer; (2) matching of grating parameters (in the toroidal variable line-space grating, the ruling parameters and the toroid radii-of-curvature have to be very well matched to correct image astigmatism); and (3) deformation of the primary mirror and/or grating due to mounting stress.

Possibly linked to this issue, we noticed a systematic bias in measurements of Doppler velocities correlated to the intensity gradients -- an effect qualitatively similar to what was reported in CDS \citep{Haugan:1999} and Hinode/EIS \citep{Young:2012}, although with a larger magnitude. The source of this bias appears to be a combination of anisotropic PSFs (i.e. astigmatism) in both the telescope and spectrometer sections. An effort is on-going to model the effect and to devise corrective actions. As of today, we recommend not to interpret Doppler velocities in SPICE data without contacting the instrument team for advice.

Initial in-flight testing of the instrument raster field of view shows that the full scan range of 16\arcmin\ will be available at distances of 0.5~AU and closer to the Sun.  As expected prior to launch, at larger distances the cooler instrument temperatures reduce the scan range.  At 1~AU, the range is 12.8\arcmin\ (80\% of maximum).  A full characterisation of the available range versus distance is on-going during the Solar Orbiter cruise phase.

\end{document}